\documentclass[12pt]{article}
\usepackage{graphicx}

\newcommand{\beq}{\begin{equation}}
\newcommand{\eeq}{\end{equation}}
\newcommand{\beqn}{\begin{eqnarray}}
\newcommand{\eeqn}{\end{eqnarray}}
\newcommand{\beqns}{\begin{eqnarray*}}
\newcommand{\eeqns}{\end{eqnarray*}}

\begin{document}
\begin{titlepage}
\begin{center}

\hfill USTC-ICTS-07-22\\
\hfill October 2007\\

\vspace{2.5cm}

{\Large {\bf  A note on the mass splitting of $K^*(892)$}} \vspace*{1.0cm}\\
{  Dao-Neng Gao$^\dagger$ and Mu-Lin Yan$^\ddagger$} \vspace*{0.3cm} \\
{\it\small Interdisciplinary Center for Theoretical Study and
Department of Modern Physics,\\ University of Science and
Technology of China, Hefei, Anhui 230026 China}\vspace*{1cm}
\end{center}
\begin{abstract}
\noindent Belle Collaboration reported a new observed value of
$K^{*-}(892)$ mass by studying $\tau^-\to K_S\pi^- \nu_\tau$
decay, which is significantly different from the current world
average value given by Particle Data Group 2006. Motivated by this
new data, we revisit the issue on the $K^{*0}(892)-K^{*\pm}(892)$
mass splitting. Our theoretical estimation favors the new
measurement by Belle Collaboration. Therefore further experimental
efforts are urgently needed to improve our understanding of these
issues.

\end{abstract}

\vfill \noindent

$^{\dagger}$ E-mail: ~gaodn@ustc.edu.cn

$^{\ddagger}$ E-mail: ~mlyan@ustc.edu.cn
\end{titlepage}

\section{Introduction}
The problem of the mass splitting between the neutral $K^*(892)$
(i.e. $K^{*0}$ or $\bar{K}^{*0}$) and the charged $K^{*}(892)$
(i.e. $K^{*\pm}$) has a long history. Experimentally, given by
Particle Data Group 2006 \cite{PDG06}, one has \beq\label{K*ms1}
m_{K^{*0}}-m_{K^{*\pm}}=6.7\pm 1.2 ~{\rm MeV}. \eeq On the other
hand, using the observed values of $K^{*0}$ and $K^{*\pm}$ masses
shown in Ref. \cite{PDG06}, \beqn\label{K*mass}
m_{K^{*0}}=896.00\pm0.25~{\rm MeV},\\m_{K^{*\pm}}=891.66\pm
0.26~{\rm MeV}, \eeqn we obtain
\beq\label{K*ms2}m_{K^{*0}}-m_{K^{*\pm}}=4.34\pm 0.36 ~{\rm
MeV},\eeq
 with smaller central value and less uncertainty by contrast with eq. (\ref{K*ms1}).
As a conservative estimation, we may thus obtain the experimental
value of the mass splitting of $K^{*}$-mesons, denoted by ``expt",
as \beq\label{K*ms3} (m_{K^{*0}}-m_{K^{*\pm}})_{\rm expt}\sim 4 ~
{\rm to}~8 ~{\rm MeV}. \eeq

It is well known that, for SU(3) flavor multiplets of hadrons, the
mass splittings between their isospin components are caused by two
effects: ~(i) $m_u\neq m_d$ (inequality of $u$-$d$ quark masses);
~(ii) the electromagnetic interactions inside hadrons.
Consequently, the observed mass splitting of $K^*$-mesons can be
expressed as follows
\beq\label{K*ms4}(m_{K^{*0}}-m_{K^{*\pm}})_{\rm
expt}=(m_{K^{*0}}-m_{K^{*\pm}})_{\rm
QM}+(m_{K^{*0}}-m_{K^{*\pm}})_{\rm EM}, \eeq where the subscript
QM denotes the contribution due to $u$-$d$ quark mass difference;
EM denotes the one due to electromagnetic interaction, so the
second term on the right side of the above equation is called
EM-mass difference. Note that these two parts contributions (QM
and EM) cannot be measured directly, theoretical calculations for
them are therefore necessary.

Usually, the EM-masses of neutral hadrons are smaller than ones of
their charged partners. For instance, the EM-masses of neutron,
$\pi^0$, and $K^0(\bar{K}^0)$ are smaller than the EM-masses of
proton, $\pi^\pm$, and $K^\pm$ \cite{LYL86, Das67, GLY97, GY00},
respectively. Thus it is reasonable to assume that
\beq\label{K*EM0} (m_{K^{*0}}-m_{K^{*\pm}})_{\rm EM}< 0,\eeq which
leads to \beq\label{K*ms5} (m_{K^{*0}}-m_{K^{*\pm}})_{\rm
expt}<(m_{K^{*0}}-m_{K^{*\pm}})_{\rm QM}.\eeq This implies that,
from eq. (\ref{K*ms3}), a relative large contribution to
$(m_{K^{*0}}-m_{K^{*\pm}})_{\rm QM}$ is required.

Very recently, Belle Collaboration reported a new measurement of
$K^{*-}(892)$ mass by studying $\tau^-\to K_S\pi^-\nu_\tau$ decay
\cite{Belle07} \beq\label{K*-mass} m_{K^{*-}}=895.47\pm 0.20 ({\rm
stat.})\pm 0.44({\rm syst.})\pm 0.59 ({\rm mod.})~{\rm MeV},\eeq
which is significantly different from the current world average
value in \cite{PDG06}. This will give (here we assume that the
neutral $K^*$ mass remains unchanged), \beq\label{K*ms6}
(m_{K^{*0}}-m_{K^{*\pm}})_{\rm expt}=0.53\pm 0.80 ~{\rm MeV},\eeq
which is quite a small value. If the Belle data is confirmed,
there would obviously exist some discrepancy between (\ref{K*ms6})
and (\ref{K*ms3}). In order to clarify this discrepancy,
experimentally, it is urgent to confirm or rule out the Belle
result (\ref{K*-mass}); and also it is important to carry out a
new measurement of the neutral $K^*$ mass. Meanwhile, further
theoretical investigations on this issue will be very helpful.

Unfortunately, it is still an open question to calculate the mass
splittings of the low-lying mesons from the first principle of
quantum chromodynamics (QCD) due to the non-perturbative feature
of QCD. Therefore, at the present stage, one generally appeals to
the low energy effective models inspired by QCD. For our purpose,
in order to get a consistent evaluation, then to understand the
above possible discrepancy, one should adopt the theoretical
framework in which both $(m_{K^{*0}}-m_{K^{*\pm}})_{\rm QM}$ and
$(m_{K^{*0}}-m_{K^{*\pm}})_{\rm EM}$ can be computed
systematically. Our previous studies \cite{GY98, GLY97}, in the
framework of chiral constituent quark model, have done such job
already. Actually our analysis \cite{GY98} favors a small value of
the $K^*(892)$ mass splitting, which is consistent with eq.
(\ref{K*ms6}). The purpose of the present note is to make this
point more clear.

\vskip0.8cm \section{Model estimation} Chiral constituent quark
model (ChQM) is developed by Manohar and Georgi \cite{MG84}, and
the vector meson ($\omega$-meson) is firstly introduced into this
model in Ref. \cite{LYL91} to study quark spin contents using the
chiral soliton approach. The author of Ref. \cite{Li95} further
extended it by including the low-lying 1$^-$(vector) and
1$^+$(axial-vector) mesons. This model has been investigated
extensively \cite{GLY97,GY98,Li97,LGY98,ZWY00,WY00,LW02} and its
theoretical results agree well with the data. The electromagnetic
interaction of mesons in ChQM has been well established via vector
meson dominance, which makes it possible to evaluate the EM-masses
of low-lying mesons in this framework \cite{GLY97}.

In Ref. \cite{GY98}, the mass splittings of vector mesons
generated from the quark mass effect in ChQM have been derived at
the leading order in quark mass expansion, and the explicit mass
formulae for the $K^*$-mesons have been obtained, as shown in eqs.
(19) and (20) of the paper. By making a reasonable approximation
in numerology, it has been found that
\beq\label{K*ms7}(m_{K^{*0}}-m_{K^{*\pm}})_{\rm
QM}=\frac{m_V}{4\pi^2 g^2}\frac{m_d-m_u}{m}\simeq
\frac{1}{2}(m_d-m_u). \eeq  As pointed out in Ref. \cite{Li95}, a
typical scale or cutoff  $\Lambda$, which is reflected by an
intrinsic parameter $g$ of the model, has been introduced. The
value of the quark mass parameters, which is actually scale
dependent, should thus be evaluated at this scale. From
$\rho^0-\omega$ mixing, the author of Ref. \cite{GY98} further
determined (similar studies have been done in \cite{WY00, LW02})
\beq\label{qmmd}m_d-m_u= 6.14\pm 0.36 ~{\rm MeV}.\eeq
Consequently, we get \cite{GY98, Yan00} \beq\label{K*MS8}
(m_{K^{*0}}-m_{K^{*\pm}})_{\rm QM}=3.07\pm 0.18 ~{\rm MeV}.\eeq
This indicates inequality (\ref{K*ms5}) may not hold for
(\ref{K*ms3}); however, it is of no problem for (\ref{K*ms6}). It
will be shown below that, when we include the EM-masses
contributions, the situation for (\ref{K*ms3}) will be much worse.

In Ref. \cite{GLY97}, electromagnetic mass splittings of $\pi$,
$K$, $a_1$, $K_1$, and $K^*(892)$ have been calculated to one-loop
order and $O(\alpha_{\rm EM})$, which gives
\beq\label{K*emms}(m_{K^{*0}}-m_{K^{*\pm}})_{\rm EM}=-1.76~{\rm
MeV}.\eeq Here we have corrected a sign error for the EM-masses of
the vector and axial-vector mesons obtained in \cite{GLY97} (note
that there is no sign error in the case of pseudoscalar mesons),
which has been firstly pointed out in \cite{GY02}.

Now from eqs. (\ref{K*MS8}) and (\ref{K*emms}), we get our
estimation in ChQM for the mass splitting of $K^*(892)$
\beqn\label{K*msth}(m_{K^{*0}}-m_{K^{*\pm}})_{\rm
theory}&=&(m_{K^{*0}}-m_{K^{*\pm}})_{\rm
QM}+(m_{K^{*0}}-m_{K^{*\pm}})_{\rm EM}\nonumber\\&\simeq& 1.3~{\rm
MeV}.\eeqn This is consistent with eq. (\ref{K*ms6}), in which the
new data by Belle Collaboration has been used; however, is
inconsistent with eq. (\ref{K*ms3}) estimated from the current
world average values.

 \section{Discussions and remarks}
Motivated by the new measurement of $K^{*-}(892)$ mass reported by
Belle Collaboration, we reexamine the mass splitting between the
neutral $K^*(892)$ and the charged $K^{*}(892)$. Our analysis
shows that there might exist some discrepancy between this new
result and the corresponding world average value by Particle Data
Group 2006 if the mass of $K^{*0}(892)$ could keep unchanged. In
the framework of ChQM, we give a theoretical estimation as
$(m_{K^{*0}}-m_{K^{*\pm}})_{\rm theory}=1\sim 2 ~{\rm MeV}$, which
seems to support the Belle data.

It has been pointed out in \cite{Belle07} that, none of the
previous mass measurements of the charged $K^*(892)$ listed in
\cite{PDG06}, all of which were performed more than twenty years
ago, present the systematic uncertainties for their measurements;
more importantly, all those earlier mass measurements listed there
come from analysis of hadronic reactions and include the effects
of final state interaction while Belle Collaboration presents the
measurement based on $\tau^-$ decays, where the decay products of
the $K^{*-}(892)$ are the only hadrons involved.

For the neutral $K^*$ masses, the situation in \cite{PDG06} is
very different from the charged case (for $K^{*\pm}$, only old
data obtained more than twenty years ago are adopted there). In
2005 FOCUS Collaboration has reported a measurement of the
$K^{*0}$ masses as \cite{FOCUS05}
\beq\label{Focus05}m_{K^{*0}}=895.41\pm0.32^{+0.35}_{-0.43}~{\rm
MeV}\eeq by studying the semileptonic $D^+\to K^-\pi^+\mu^+\nu$
decay. Similarly to the case of $\tau^-\to K_S\pi^-\nu_\tau$ decay
for the $K^{*-}$ mass measurement, this semileptonic $D^+$ decay
could also provide a nice place to study $K\pi$ system in the
absence of interactions with other hadrons, in which $K^{*0}$
mesons are the only hadrons produced in the decay final state. It
is noteworthy that the recent FOCUS measurement (\ref{Focus05}) is
close to the world average value of the $K^{*0}(892)$ mass in
\cite{PDG06}. This indicates that the recent data prefer to the
small value of the mass splitting $m_{K^{*0}}-m_{K^{*\pm}}$, which
is consistent with our model estimation; however, contradicts to
the world average values by Particle Data Group 2006. Explicitly,
if we only consider the values in eqs. (\ref{K*-mass}) and
(\ref{Focus05}), which are the most recent data for the $K^{*-}$
and $K^{*0}$ masses, respectively, one has
\beq\label{K*ms9}(m_{K^{*0}}-m_{K^{*\pm}})_{\rm
FOCUS-Belle}=-0.06\pm 0.91 ~{\rm MeV}. \eeq

Another possible interesting and precise experiment has been
proposed in \cite{Yan00,Sun04}
 that $K^*(892)$ masses can be measured in BES at BEPC, especially for the neutral $K^*$ mass.
 Since BESII at BEPC has collected about $5.77\times 10^7$
$J/\Psi$ events, it is practicable to take $J/\Psi$ as the source
of $K^*(892)$. The branching ratio for $J/\Psi\to \bar{K}^0
K^{*0}$ is $4.2\times 10^{-3}$; for $J/\Psi\to K\bar{K}\pi$,
$6\times 10^{-3}$; for $K^*\to K\pi$, about 100\%. Therefore,
studying three-body decay processes $J/\Psi\to K \bar{K}\pi$, one
can determine the location of the resonance of $K^\mp\pi^\pm$(i.e.
$K^{*0}$ or $\bar{K}^{*0}$), and measure the neutral $K^{*}$ mass
with the error below $1$ MeV. More accurate measurements can be
expected in BESIII after the BESII detector is upgraded.

Our theoretical calculation (\ref{K*msth}) is not a model
independent estimation. However, the discrepancy between the most
recent data given by eq. (\ref{K*ms6}) or (\ref{K*ms9}) and the
current world average value by Particle Data Group 2006 does
exist. Future dedicated measurements of the $K^*(892)$ (including
both $K^{*0}$ and $K^{*\pm}$) masses with high precision are
necessary to clarify this discrepancy. We therefore urge our
experimental colleagues to produce more data in order to get a
solid and more meaningful conclusion.

\section*{Acknowledgements}
We are grateful to Bo-Qiang Ma, Kim Maltman, and Han-Qing Zheng
for very helpful discussions. This work is supported in part by
National Natural Science Foundation of China under Grant No.
10275059, 10775124 and 90403021, and KJCX2-SW-N10 of the Chinese
Academy of Sciences.

\end{document}